\documentclass[a4paper,12pt]{article}
\usepackage{epsfig}
\usepackage{latexsym,amssymb}
\usepackage{graphics}
\usepackage{pstricks,psfrag}
\usepackage{amsmath}
\usepackage{subfigure}

\makeatletter
\def\vereq#1#2{
\lower3pt\vbox{\baselineskip1.5pt \lineskip1.5pt
\ialign{$\m@th#1\hfill##\hfil$\crcr#2\crcr\sim\crcr}}}
\makeatother

\begin{document}

\begin{titlepage}
\begin{center}
\hfill    CERN-PH-TH/2011-294\\

\vskip 1cm

{\large \bf Invariants, Alignment and the Pattern of Fermion Masses and Mixing}

\vskip 1cm

G.C. Branco$^{1,2}$\footnote{E-mail: gbranco@ist.utl.pt}
and  
J.I. Silva-Marcos$^{2}$\footnote{E-mail: juca@cftp.ist.utl.pt}

\vskip 0.07in

{\em $^{1}$CERN Theory Division,\\ CH-1211 Geneva 23, Switzerland\\

\vskip 0.07in

$^2$CFTP, Departamento de F\'{\i}sica,\\}
{\it  Instituto Superior T\'ecnico, Universidade T\'ecnica de Lisboa, }
\\
{\it
Avenida Rovisco Pais nr. 1, 1049-001 Lisboa, Portugal}
\end{center}

\vskip 3cm

\begin{abstract}
We show that the
main features of the pattern of 
fermion masses and mixing can be expressed
in terms of simple relations among weak-basis invariants.
In the quark
sector, we identify the weak-basis invariants which signal the observed
alignment of the
up and down quark mass matrices in flavour space. 
In the lepton sector, we indicate how a set of conditions on weak-basis invariants
can lead to an approximate tribimaximal lepton mixing matrix.
We also show the usefulness of 
these invariants in the study of specific
ans\"{a}tze for the flavour structure of fermion
mass matrices.
\end{abstract}

\end{titlepage}

\newpage

\section{\protect\bigskip Introduction}

In the past few years there has been a remarkable progress in the
determination of fermion masses and mixing \cite{mix}, involving advances
both in theory and experiment. In the quark sector, input from experiment
includes the knowledge of $|V_{us}|$, $|V_{cb}|$, $|V_{ub}/V_{cb}|$, $%
|V_{td}| $, $|V_{ts}|$, together with the measurement of the rephasing
invariant angles $\beta $ and $\gamma $. In the framework of the Standard
Model (SM), these results constrain the location of the vertex of the
unitarity triangle to a small region. The measurement of the angle $\gamma $
is specially important since it provides clear evidence that the Cabibbo-
Kobayashi-Maskawa (CKM) matrix \cite{ckm} is complex, even if one allows for
the presence of New Physics beyond the SM \cite{bsm}. In the leptonic
sector, non-vanishing neutrino masses and mixing has been established \cite
{neut0}. However, there are still important open questions like the 
nature
of the neutrino mass spectrum (normal hierarchy, inverted hierarchy or quasi
degenerate), the determination of whether neutrinos are Majorana or Dirac
particles, as well as the search for leptonic CP violation. In this respect,
the recent evidence in favour of a non-vanishing value of $U_{e3}$, provides
the hope of discovering leptonic CP violation in neutrino oscillations. It
is well known that having a non-vanishing $U_{e3}$\ is a necessary
requirement in order to have leptonic Dirac-type CP violation, which is
detectable in neutrino oscillations.

In spite of these developments, one does not have yet a standard theory of
flavour. One may adopt a bottom-up approach and try to discover a symmetry
principle from the observed pattern of fermion masses and mixing. One of the
difficulties that one encounters in following this approach, stems from the
fact that there is a large redundancy in the Yukawa couplings $Y_{u}$, $%
Y_{d} $\ \ which generate the quark masses $M_{u}$, $M_{d}$. One can make
weak-basis (WB) transformations which change $M_{u}$, $M_{d}$, but do not
alter their physical content. The above considerations also apply to flavour
models which postulate the existence of so-called "texture-zeros". It is
clear that these zeros only exist in a particular WB.

The above redundancy in Yukawa couplings and fermion mass matrices motivates
the use of WB invariants, i.e. functions of quark masses which do not change
when one preforms a WB transformation. These WB invariants are very useful
in the analysis of CP violation, where they have been derived from first
principles \cite{principles} and have been applied to both the quark \cite
{cp-quark} and lepton \cite{cp-lepton} sectors, including leptogenesis, as
well as to the Higgs sector \cite{cp-higgs}.

In this paper, we show that the main features of the pattern of fermion
masses and mixing can be expressed in terms of simple relations involving
only WB invariants. We introduce the concept of "alignment", which can be
understood in the following way. For definiteness, let us consider the quark
sector, where small flavour mixing is indicated by experiment. Small mixing
implies that there is a WB where both $M_{u}$ and $M_{d}$ are close to the
diagonal form. However, experiment shows more than that, it tells us that
the quark mass matrices are aligned in flavour space, meaning that there is
a basis where $M_{d}=diag(m_{d},m_{s},m_{b})$ and $M_{u}$ is close to $%
diag(m_{u},m_{c},m_{t})$ and not to $diag(m_{t},m_{c},m_{u})$, for example.
Obviously, only the relative ordering of the eigenvalues of $M_{u}$, $M_{d}$
is physically meaningful, since by making a WB transformation, one can
change simultaneously the eigenvalue ordering in $M_{u}$, $M_{d}$.

At this point, it is worth recalling that in the context of the SM, the
Yukawa couplings leading to $M_{u}$ and $M_{d}$ are entirely independent,
there is no "dialog". between $Y_{u}$ and $Y_{d}$. Therefore, in the context
of the SM, or its minimal supersymmetric extension, alignment is in no way
more "natural" than misalignment. Quite on the contrary, if one 
considers
the manifold of matrices $M_{u}$, $M_{d}$ leading to "small mixing" as
previously defined, the probability of having alignment is only 1/6. 

In this
paper we will show how WB invariants can distinguish not only between small
mixing and large mixing, but also between alignment and misalignment.

The paper is organized as follows:. In the next section, we consider the
quark sector, where we illustrate the usefulness of WB invariants for two
and three generations. In section 3, we apply WB invariants to the study of
some ans\"{a}tze for quark mass matrices, while in section 4 we briefly
study the lepton sector, showing in particular how the observed pattern of
leptonic mixing can be obtained through a set of conditions on WB
invariants. Our conclusions are contained in section 5.

\section{The quark sector}

\subsection{Two quark generations}

For simplicity and in order to explain the concept of alignment and its
connection to weak-basis invariants, we consider first the case of two
generations. The three important features of quark masses and mixing are:

\begin{description}
\item  (i) Hierarchical quark masses,

\item  (ii) Small mixing,

\item  (iii) Alignment.
\end{description}

We consider separately each one of these features, emphasizing that they are
logically independent, at least in the context of the SM. We shall identify
how each one of these features can be expressed in terms of weak-basis
invariants.

\textbf{(i) Hierarchical masses}

\noindent For two generations, the fact that quark masses are hierarchical
can be expressed in terms of invariants as, 
\begin{equation}
r_{1}\equiv \frac{Det[H]}{\left( \frac{1}{2}Tr[H]\right) ^{2}}\ll 1
\label{r1}
\end{equation}
where $H\equiv MM^{\dagger }$\ denotes either $H_{d}$\ or $H_{u}$. The case
of exact degeneracy corresponds to $r=1$.

\textbf{(ii) Small mixing}

\noindent It can be readily verified that the following relation holds 
\begin{equation}
Tr\left( [H_{u},H_{d}]^{2}\right) =-\frac{1}{2}\left( \Delta
_{12}^{d}\right) ^{2}\left( \Delta _{12}^{u}\right) ^{2}\sin ^{2}(2\theta )
\label{i1}
\end{equation}
where $\Delta _{12}^{d}=m_{d}^{2}-m_{s}^{2}$, $\Delta
_{12}^{u}=m_{c}^{2}-m_{u}^{2}$ and $\theta $\ denotes the Cabibbo angle. Let
us consider the following invariant ratio 
\begin{equation}
r_{2}\equiv \frac{\left| Tr\left( [H_{u},H_{d}]^{2}\right) \right| }{\frac{1%
}{2}\left( Tr[H_{d}]\right) ^{2}\left( Tr[H_{u}]\right) ^{2}}  \label{r2}
\end{equation}
Assuming that quark masses are hierarchical, which can be guaranteed through
the invariant condition of Eq. (\ref{r1}), it is clear from Eq. (\ref{i1}),
that 
\begin{equation}
r_{2}\approx \sin ^{2}(2\theta )  \label{r2s}
\end{equation}
Therefore small mixing can be achieved through the invariant condition 
\begin{equation}
r_{2}\ll 1  \label{r2ss}
\end{equation}
Maximal mixing corresponds to $\theta =45^{o}$, i.e. 
\begin{equation}
r_{2}=1  \label{r2l}
\end{equation}

\textbf{(iii) Alignment}

Small mixing means that there is a weak-basis where both $H_{d}$\ and $H_{u}$
are close to the diagonal form. As mentioned before, this is not sufficient
to have ''alignment'', since it does not guarantee the same ''ordering'' in
both $H_{d}$\ and $H_{u}$. Alignment means, of course, that in a WB where $%
H_{d}$ is close to $diag(m_{d}^{2},m_{s}^{2})$, $H_{u}$ is close to $%
diag(m_{u}^{2},m_{c}^{2})$ and not $diag(m_{c}^{2},m_{u}^{2})$. As
previously emphasized, one can change the ordering simultaneously in the up
and down sectors through a WB transformation. Only the relative ordering in
the up and down quark sectors is physically meaningful. In the case of two
generations, assuming hierarchical quark masses and small mixing, one has
alignment if for the following invariant 
\begin{equation}
\begin{array}{l}
I_{1}\equiv \frac{Tr[H_{u}]\ Tr[H_{d}]-Tr[H_{u}H_{d}]}{Tr[H_{u}]\ Tr[H_{d}]}%
=\sin ^{2}(\theta )+ \\ 
\\ 
+\left( \frac{\left( m_{d}/m_{s}\right) ^{2}}{\left( 1+\left(
m_{d}/m_{s}\right) ^{2}\right) }+\frac{\left( m_{u}/m_{c}\right) ^{2}}{%
\left( 1+\left( m_{u}/m_{c}\right) ^{2}\right) }-\frac{2\left(
m_{d}/m_{s}\right) ^{2}\left( m_{u}/m_{c}\right) ^{2}}{\left( 1+\left(
m_{d}/m_{s}\right) ^{2}\right) \left( 1+\left( m_{u}/m_{c}\right)
^{2}\right) }\right) \cos (2\theta )
\end{array}
\label{I10}
\end{equation}
the condition is satisfied: 
\begin{equation}
I_{1}\ll 1  \label{I1}
\end{equation}
On the contrary, assuming again hierarchical quark masses and small mixing,
misalignment implies: 
\begin{equation}
I_{1}\approx 1  \label{I1-a}
\end{equation}

\subsection{Three quark generations}

\subsubsection{Hierarchy of quark masses}

The hierarchy of quarks masses in both the up and down quark sectors, namely 
\begin{equation}
m_{1}^{2}\ll m_{2}^{2}\ll m_{3}^{2}  \label{ms}
\end{equation}
can be translated into invariant conditions. We introduce the Hermitian
quark mass matrices $H\equiv MM^{\dagger }$\ and the corresponding
invariants $\det (H)$, $Tr[H]$ together with the third invariant $\chi
\lbrack H]$\ which stands for $\chi \lbrack H]\equiv
m_{1}^{2}m_{2}^{2}+m_{1}^{2}m_{3}^{2}+m_{2}^{2}m_{3}^{2}$. Note that for an
Hermitian $3\times 3$ matrix $H$, one has 
\begin{equation}
\chi \lbrack H]=\frac{1}{2}\left( \left( Tr[H]\right) ^{2}-Tr[H^{2}]\right)
\label{ki1}
\end{equation}

The following invariant condition 
\begin{equation}
R_{1}\equiv \frac{\chi \lbrack H]}{Tr[H]^{2}}\ll 1  \label{hier2}
\end{equation}
implies that one of the eigenvalues of $H$ is much larger that the other
two. Finally, it can be readily verified that the condition 
\begin{equation}
R_{2}\equiv \frac{Tr[H]\ Det[H]}{\left( \chi \lbrack H]\right) ^{2}}\ll 1
\label{hier3}
\end{equation}
together with Eq. (\ref{hier2}) implies that of the two smaller eigenvalues,
one of them is much smaller than the other one, i.e. $m_{1}^{2}\ll m_{2}^{2}.
$

\subsubsection{Invariants and the pattern of mixing}

\bigskip Previously \cite{previously}, invariants were used to study
specific ans\"{a}tze where the quark mass matrices were written in an
Hermitian basis. Here, we consider WB invariants which can be applied in an
arbitrary basis, not necessarily Hermitian.

It is convenient to introduce the following dimensionless matrices with unit
trace, $Tr[h_{u,d}]=1$: 
\begin{equation}
h_{u}=\frac{H_{u}}{Tr[H_{u}]}\qquad ;\qquad h_{d}=\frac{H_{d}}{Tr[H_{d}]}
\label{tr1}
\end{equation}
and their difference: 
\begin{equation}
A\equiv h_{d}-h_{u}  \label{diffh}
\end{equation}
By construction, one has $Tr[A]=0$, which in turn implies 
\begin{equation}
\chi \left[ A\right] =-\frac{1}{2}Tr[A^{2}]  \label{chia}
\end{equation}
There is a relation between $\chi (A)$ and $I_{1}$, defined in Eq. 
(\ref{I10}%
). Indeed, from Eqs. (\ref{I10}, \ref{tr1}), one obtains 
\begin{equation}
I_{1}=1-Tr[h_{u}h_{d}]  \label{reli1}
\end{equation}
while Eqs. (\ref{diffh}, \ref{chia}) lead to 
\begin{equation}
\chi \left[ A\right] =Tr[h_{u}h_{d}]-\frac{1}{2}Tr[h_{u}^{2}]-\frac{1}{2}%
Tr[h_{d}^{2}]  \label{chia1}
\end{equation}
From Eqs. (\ref{reli1}, \ref{chia1}), one therefore gets 
\begin{equation}
\chi \left[ A\right] =1-I_{1}-\frac{1}{2}Tr[h_{u}^{2}]-\frac{1}{2}%
Tr[h_{d}^{2}]  \label{reli2}
\end{equation}
Assuming hierarchy of the quark masses, which can be implemented through the
invariants of Eqs. (\ref{hier2}, \ref{hier3}), one obtains 
\begin{equation}
\begin{array}{l}
Tr[h_{d}^{2}]=1-2\left( \frac{m_{s}}{m_{b}}\right) ^{2}+O\left( \left( \frac{%
m_{s}}{m_{b}}\right) ^{4}\right) \\ 
\\ 
Tr[h_{u}^{2}]=1-2\left( \frac{m_{c}}{m_{t}}\right) ^{2}+O\left( \left( \frac{%
m_{c}}{m_{t}}\right) ^{4}\right)
\end{array}
\label{trhud}
\end{equation}
On the other hand, an explicit evaluation of $I_{1}$ for three generations
in terms of $\left| V_{ij}\right| $ and quark mass ratios gives 
\begin{equation}
I_{1}=\left| V_{23}\right| ^{2}+\left| V_{13}\right| ^{2}+\left( \frac{m_{s}%
}{m_{b}}\right) ^{2}+\left( \frac{m_{c}}{m_{t}}\right) ^{2}+O\left( \left( 
\frac{m_{s}}{m_{b}}\right) ^{4}\right)  \label{i10}
\end{equation}
Using Eqs. (\ref{reli2}, \ref{trhud}, \ref{i10}), one finally gets: 
\begin{equation}
\left|\chi \left[ A\right]\right| =\left| V_{23}\right| ^{2}+\left|
V_{13}\right| ^{2}+O\left( \left( \frac{m_{s}}{m_{b}}\right) ^{4}\right)
\label{chia3}
\end{equation}
The usefulness of $\chi \left[ A\right] $\ is clear from Eq. (\ref{chia3}):
it gives to an excellent approximation the value of $\left| V_{23}\right|
^{2}+\left| V_{13}\right| ^{2}$.

At this stage, it is worth recalling that one of the main features of quark
mixing, is the fact that the 3rd generation almost decouples from the other
two. The deviation of exact decoupling is given by the size of $\left\vert
V_{23}\right\vert ^{2}+\left\vert V_{13}\right\vert ^{2}$. The experimental
measurement of $\left\vert V_{23}\right\vert $ and $\left\vert
V_{13}\right\vert $ shows that: 
\begin{equation}
\left\{ \left\vert V_{23}\right\vert ^{2}+\left\vert V_{13}\right\vert
^{2}\right\} ^{\exp .}=O\left( \left( \frac{m_{s}}{m_{b}}\right) ^{2}\right)
\label{expv23}
\end{equation}
This input from experiment, can be written in terms of a simple relation
among WB invariants 
\begin{equation}
\chi \left[ A\right] =O\left( \frac{\chi \left[ H_{d}\right] }{\left(
Tr[H_{d}]\right) ^{2}}\right) =O\left( \chi \left[ h_{d}\right] \right)
\label{expchia}
\end{equation}

It is worth emphasizing that, for three generations, $\chi \left[ A\right] $
is also a measure of the alignment of the down and up quark mass matrices.
Working in a WB where the up quarks are diagonal, one can choose, without
loss of generality, to order the up quarks in such a way that $%
H_{u}=diag(m_{u}^{2},m_{c}^{2},m_{t}^{2})$. In the context of small mixing,
alignment means that, in the above basis, $H_{d}$ is close to $%
diag(m_{d}^{2},m_{s}^{2},m_{b}^{2})$. In this case, $\chi \left[ A\right] $
is small. In fact, if we take the limit $m_{t}\rightarrow \infty $, $%
m_{b}\rightarrow \infty $, one has $h_{u}=diag(0,0,1)$, $h_{d}=diag(0,0,1)$
and $A$ vanishes, so $\chi \left[ A\right] =0$. On the other hand, if there
is small mixing, but no alignment, in the WB where $%
H_{u}=diag(m_{u}^{2},m_{c}^{2},m_{t}^{2})$, one may have $H_{d}$ is close to 
$diag(m_{b}^{2},m_{s}^{2},m_{d}^{2})$. In this case, $\chi \left[ A\right] $
is large. Indeed in the limit $m_{t}\rightarrow \infty $, $m_{b}\rightarrow
\infty $, one has, for this case, $h_{u}=diag(0,0,1)$ but $h_{d}=diag(1,0,0)$%
, which leads to $\left|\chi \left[ A\right]\right| =1$, signalling 
total 
misalignment.

Next, we address the question of how to use invariants to constrain
separately $\left| V_{23}\right| ^{2}$ and $\left| V_{13}\right| ^{2}$. This
is a more difficult task, involving more complicated invariants, as it was
to be expected. In order to constrain $\left| V_{13}\right| $, let us
consider the following WB invariant: 
\begin{equation}
I_{2}\equiv 1-\frac{Tr[H_{u}]\ Tr[H_{u}H_{d}]-Tr[H_{u}^{2}H_{d}]}{\chi
[H_{u}]\ Tr[H_{d}]}  \label{i2}
\end{equation}
This invariant can be readily calculated and one obtains in the chiral
limit, i.e. when $m_{u},m_{d}=0$: 
\begin{equation}
I_{2}=\frac{\frac{m_{s}^{2}}{m_{b}^{2}}\ |V_{12}|^{2}+|V_{13}|^{2}}{1+\frac{%
m_{s}^{2}}{m_{b}^{2}}}  \label{i2-limit}
\end{equation}
It is clear from Eq. (\ref{i2-limit}) that if we constrain $I_{2}$ to be of
order $\lambda ^{6}$, $\lambda $\ denoting the Cabibbo angle, then $|V_{13}|$%
\ is at most of order $\lambda ^{3}$. It can be shown that this conclusion
holds when one does not assume the chiral limit. Indeed, an exact
calculation of \ gives: 
\begin{equation}
\begin{array}{l}
I_{2}=\frac{1}{\left[ 1+\left( \frac{m_{s}}{m_{b}}\right) ^{2}+\left( \frac{%
m_{d}}{m_{b}}\right) ^{2}\right] \left[ 1+\left( \frac{m_{u}}{m_{c}}\right)
^{2}+\left( \frac{m_{u}}{m_{t}}\right) ^{2}\right] }\ \cdot \\ 
\\ 
(\ |V_{13}|^{2}+\left( \frac{m_{u}}{m_{c}}\right) ^{2}|V_{23}|^{2}+\left( 
\frac{m_{u}}{m_{t}}\right) ^{2}|V_{33}|^{2}+ \\ 
\\ 
+\left( \frac{m_{s}}{m_{b}}\right) ^{2}\left[ |V_{12}|^{2}+\left( \frac{m_{u}%
}{m_{c}}\right) ^{2}|V_{22}|^{2}+\left( \frac{m_{u}}{m_{t}}\right)
^{2}|V_{32}|^{2}\right] + \\ 
\\ 
+\left( \frac{m_{d}}{m_{b}}\right) ^{2}\left[ |V_{11}|^{2}+\left( \frac{m_{u}%
}{m_{c}}\right) ^{2}|V_{21}|^{2}+\left( \frac{m_{u}}{m_{t}}\right)
^{2}|V_{31}|^{2}\right] \ )
\end{array}
\label{i2-exact}
\end{equation}
From Eq. (\ref{i2-exact}), and given the quark mass hierarchy, one concludes
that putting $I_{2}\approx \lambda ^{6}$ constrains $|V_{13}|$ to be at most
of order $\lambda ^{3}$. Then, from Eq. (\ref{chia3}), it follows that
setting $\chi [A]\approx \lambda ^{4}$ constrains $|V_{23}|$\ to be of order 
$\lambda ^{2}$, as indicated by experiment. We have thus shown how to fix
separately $|V_{23}|$ and $|V_{13}|$ through WB invariants.

In order to constrain\ $|V_{12}|$, it is convenient to use WB invariants
involving $H_{u,d}^{-1}$. Let us define 
\begin{equation}
\widehat{A}=\widehat{h}_{d}-\widehat{h}_{u}  \label{diffinv}
\end{equation}
where 
\begin{equation}
\widehat{h}_{u}=\frac{H_{u}^{-1}}{Tr[H_{u}^{-1}]}\qquad ;\qquad \widehat{h}%
_{d}=\frac{H_{d}^{-1}}{Tr[H_{d}^{-1}]}  \label{hinv}
\end{equation}
We have assumed that none of the quark masses vanish, as indicated by
experiment and theory. In the weak basis where the up quark mass matrix is
diagonal, we have 
\begin{equation}
\begin{array}{l}
\widehat{h}_{u}=\frac{1}{\left( 1+\frac{m_{u}^{2}}{m_{c}^{2}}+\frac{m_{u}^{2}%
}{m_{t}^{2}}\right) }\ diag\left( 1,\frac{m_{u}^{2}}{m_{c}^{2}},\frac{%
m_{u}^{2}}{m_{t}^{2}}\right) \\ 
\\ 
\widehat{h}_{d}=\frac{1}{\left( 1+\frac{m_{d}^{2}}{m_{s}^{2}}+\frac{m_{d}^{2}%
}{m_{b}^{2}}\right) }\ V\cdot diag\left( 1,\frac{m_{d}^{2}}{m_{s}^{2}},\frac{%
m_{d}^{2}}{m_{b}^{2}}\right) \cdot V^{\dagger }
\end{array}
\label{hhoed}
\end{equation}
Note the eigenvalues of $\widehat{h}_{u,d}$, denoted by $\widehat{\mu }_{i}$
satisfy an inverted hierarchy: $\widehat{\mu }_{1}\gg \widehat{\mu }_{2}\gg 
\widehat{\mu }_{3}$. We evaluate now $\chi \lbrack \widehat{A}]$, obtaining: 
\begin{equation}
\left\vert \chi (\widehat{A})\right\vert =\left\vert V_{12}\right\vert
^{2}+\left\vert V_{13}\right\vert ^{2}-2\left( \frac{m_{d}}{m_{s}}\right)
^{2}\left\vert V_{12}\right\vert ^{2}+O(\lambda ^{8})  \label{ahoed}
\end{equation}
If one constrains $\chi \lbrack \widehat{A}]$ to be of order $\lambda ^{2}$,
one necessarily has $\left\vert V_{12}\right\vert \cong \lambda $, taking
into account that $\left\vert V_{13}\right\vert ^{2}$ was already
constrained to be of order $\lambda ^{6}$.

In order to complete the determination of $V^{CKM}$ through WB invariants,
we have to address the question of CP violation. It has been shown \cite
{principles} from first principles that the vanishing of the following WB
invariant is a necessary condition for CP invariance in the SM, for an
arbitrary number of generations: 
\begin{equation}
I^{CP}\equiv Tr\left( [H_{u},H_{d}]^{3}\right)  \label{icp}
\end{equation}
For three generations $I^{CP}=0$ is a necessary and sufficient condition for
CP invariance. In terms of physical quantities, one has 
\begin{equation}
I^{CP}=G\ \mathrm{{Im}[V_{12}V_{23}V_{13}^{*}V_{22}^{*}]}  \label{icp0}
\end{equation}
where $
G=6i(m_{b}^{2}-m_{s}^{2})(m_{b}^{2}-m_{d}^{2})(m_{s}^{2}-m_{d}^{2})(m_{t}^{2}-m_{c}^{2})(m_{t}^{2}-m_{u}^{2})(m_{c}^{2}-m_{u}^{2}) 
$. If we now set $\frac{1}{G}I^{CP}$ to be of order $\lambda ^{6}$, and take
into account that $\mathrm{Im}[V_{12}V_{23}V_{13}^{*}V_{22}^{*}]=$ $\left|
V_{12}\right| \left| V_{23}\right| \left| V_{13}\right| \left| V_{22}\right|
\sin (\phi )$, (with $\phi =\arg [V_{12}V_{23}V_{13}^{*}V_{22}^{*}]$), then
we conclude that this constrains $\sin (\phi )$ to be of order one.

We have thus shown, without going through any process of diagonalization of
the quark mass matrices, how a set of WB invariants can completely fix the
pattern of mixing and strength of CP violation present in $V^{CKM}$.

\section{Applying Invariants to various ans\"{a}tze}

\subsection{General remark}

Next, we show the usefulness of the WB invariants introduced in the previous
section and apply these invariants to some specific ans\"{a}tze.

First, we derive some general results which apply to any flavour model where
both the up and down Hermitian squared quark mass matrices with trace
normalized to unity are equal to some fixed matrix $\Delta _{o}$ of order
one plus some small perturbation denoted by $\left( \varepsilon A\right)
_{u,d}$: 
\begin{equation}
h_{d}=\Delta _{o}+\varepsilon _{d}A_{d}\qquad ;\qquad h_{u}=\Delta
_{o}+\varepsilon _{u}A_{u}  \label{small}
\end{equation}
An example could be the case where $\Delta _{o}$ stands for the so-called
democratic matrix, where all elements are equal, but our results apply to a
broader class of flavour matrices. It is clear that Eq. (\ref{small}) is a
sufficient condition to obtain alignment, since it follows from Eq. (\ref
{small}), that $A=h_{d}-h_{u}=\varepsilon _{d}A_{d}-\varepsilon _{u}A_{u}$
is small. Let us consider now those ans\"{a}tze where the following further
conditions are satisfied 
\begin{equation}
\begin{array}{l}
\left| \varepsilon _{u}\right| \ll \left| \varepsilon _{d}\right| \qquad
;\qquad Tr[A]_{u,d}=0\qquad ;\qquad \left( Tr[\Delta _{o}]\right)
^{2}=Tr[\Delta _{o}^{2}] \\ 
\\ 
Tr[A_{u,d}\ \Delta _{o}]\leq O(\ \varepsilon)_{u,d}\qquad
\end{array}
\label{valid}
\end{equation}
It follows from Eq. (\ref{valid}), that to a good approximation $A\approx
\varepsilon _{d}A_{d}$ and one obtains 
\begin{equation}
\left| \chi (A)\right| =\varepsilon _{d}^{2}\ \left| \chi [A_{d}^{2}]\right|
=\frac{1}{2}\varepsilon _{d}^{2}Tr[A_{d}^{2}]=O\left( \varepsilon
_{d}^{2}\right)  \label{apchi}
\end{equation}
Furthermore, using the conditions of Eq. (\ref{valid}) and computing $\chi
(h_{d})$, one gets 
\begin{equation}
\begin{array}{l}
\left| \chi (h_{d})\right| = \\ 
\\ 
=\frac{1}{2}\left| \left( Tr[\Delta _{o}]\right) ^{2}-Tr[\Delta
_{o}^{2}]-\varepsilon _{d}Tr[A_{d}\ \Delta _{o}+\Delta
_{o}A_{d}]-\varepsilon _{d}^{2}Tr[A_{d}^{2}]\right| =O\left( \varepsilon
_{d}^{2}\right)
\end{array}
\label{trhd}
\end{equation}
Therefore, one finds 
\begin{equation}
\left| \chi (A)\right| =O\left( \ \left| \chi (h_{d})\right| \right)
=O(\left( \frac{m_{s}}{m_{b}}\right) ^{2})  \label{cases}
\end{equation}
Note that Eq. (\ref{cases}) coincides with Eq. (\ref{expchia}) and
therefore, for the whole class of ans\"{a}tze satisfying the generic
conditions of Eqs. (\ref{small}, \ref{valid}), one has the correct
prediction 
\begin{equation}
\left| V_{23}\right| ^{2}+\left| V_{13}\right| ^{2}=O(\left( \frac{m_{s}}{%
m_{b}}\right) ^{2})  \label{result0}
\end{equation}
This is a remarkable result. Using WB invariants, one can show that a whole
class of ans\"{a}tze for $M_{u}$, $M_{d}$ satisfying the generic of Eq. (\ref
{valid}), satisfies Eq. (\ref{result0}), which is one of the experimentally
observed salient features of $V^{CKM}$.

\subsection{The USY ansatz}

We now apply our invariants to the hypothesis of Universality of 
Strength of
Yukawa (USY) couplings \cite{usy}, \cite{usy1}, \cite{usy2}, where all
Yukawa couplings have equal moduli, the flavour dependence being all
contained in their phases. For definiteness, let us consider the case where $%
M_{u,d}$ have the symmetric form: 
\begin{equation}
M_{u,d}=c_{u,d}\left( 
\begin{array}{ccc}
1 & 1 & e^{i(\alpha -\beta )} \\ 
1 & 1 & e^{i(\alpha )} \\ 
e^{i(\alpha -\beta )} & e^{i(\alpha )} & e^{i(\alpha )}
\end{array}
\right) _{u,d}  \label{usy1}
\end{equation}
Computing the invariants of the associated $h_{u,d}$, 
\begin{equation}
\begin{array}{l}
Det[h]=\frac{4^{2}}{9^{3}}\sin ^{4}(\frac{\beta }{2}) \\ 
\\ 
\chi [h]=\frac{4}{9^{2}}\left[ \sin ^{2}(\frac{\alpha }{2})+4\sin ^{2}(\frac{%
\beta }{2})+\sin ^{2}(\frac{\alpha -2\beta }{2})+2\sin ^{2}(\frac{\alpha
-\beta }{2})\right]
\end{array}
\label{inv1}
\end{equation}
we find that in leading order the parameters $\alpha _{u,d}$ and $\beta
_{u,d}$ are small, 
\begin{equation}
\begin{array}{c}
\left| \alpha _{d}\right| =\frac{9}{2}\frac{m_{s}}{m_{b}}\quad ;\quad \left|
\beta _{d}\right| =3\sqrt{3}\frac{\sqrt{m_{d}m_{s}}}{m_{b}} \\ 
\\ 
\left| \alpha _{u}\right| =\frac{9}{2}\frac{m_{c}}{m_{t}}\quad ;\quad \left|
\beta _{u}\right| =3\sqrt{3}\frac{\sqrt{m_{u}m_{c}}}{m_{t}}
\end{array}
\label{ab}
\end{equation}
and that the $h_{u,d}$ computed from Eq. (\ref{usy1}) have the form: 
\begin{equation}
h_{u,d}=\frac{\Delta }{3}+\left( \varepsilon A\right) _{u,d}\quad ;\quad
\Delta =\left( 
\begin{array}{ccc}
1 & 1 & 1 \\ 
1 & 1 & 1 \\ 
1 & 1 & 1
\end{array}
\right) _{u,d}  \label{demo}
\end{equation}
where $\Delta $ is the democratic mass matrix and $\left( \varepsilon
A\right) _{u,d}$ are matrices of order $\alpha _{u,d}$ and $\beta _{u,d}$.
Up to second order in the largest parameter $\alpha $, we find 
\begin{equation}
\left( \varepsilon A\right) _{u,d}=\frac{1}{9}\left( 
\begin{array}{lll}
0 & -i\beta & -2i\alpha -\alpha ^{2} \\ 
i\beta & 0 & -2i\alpha -\alpha ^{2}+i\beta \\ 
2i\alpha -\alpha ^{2} & 2i\alpha -\alpha ^{2}-i\beta & 0
\end{array}
\right) _{u,d}  \label{eps}
\end{equation}
The form of $h_{u,d}$ corresponds to our general conditions in Eqs. 
(\ref{small}, \ref{valid}) and we find that $\left| \chi (A)\right| $ is 
indeed small. Thus,
the USY scenario implies alignment. Furthermore, we find that in leading
order 
\begin{equation}
\left| \chi (A)\right| =\left| \chi (\varepsilon _{d}A_{d})\right|
\label{chiusyd}
\end{equation}
and that 
\begin{equation}
\left| \chi (\varepsilon _{d}A_{d})\right| =2\left| \chi (h_{d})\right|
\label{chiusyd1}
\end{equation}
Therefore, combining with Eq. (\ref{chia3}), we obtain in leading order 
\begin{equation}
\left| V_{23}\right| ^{2}+\left| V_{13}\right| ^{2}=2\left( \frac{m_{s}}{%
m_{b}}\right) ^{2}  \label{usyv23}
\end{equation}

In addition, one obtains for the invariant $I_{2}$ associated with $%
\left\vert V_{13}\right\vert $ of Eq. (\ref{i2}) and in the limit $m_{u}=0$,
the exact result 
\begin{equation}
I_{2}=\frac{2}{9}\sin ^{2}\left( \frac{\beta _{d}}{2}\right)  \label{i2-usy}
\end{equation}
which combined with Eqs. (\ref{i2-limit}, \ref{ab}) leads, in leading order
to the following expression: 
\begin{equation}
\frac{m_{s}^{2}}{m_{b}^{2}}\ |V_{12}|^{2}+|V_{13}|^{2}=\frac{3}{2}\frac{%
m_{d}m_{s}}{m_{b}^{2}}  \label{i2-usy0}
\end{equation}
The results, expressed in Eqs. (\ref{usyv23}, \ref{i2-usy0}) are in
agreement with the results which were obtained for this ansatz 
\cite{usy2},
where in leading order, it was found that $\left\vert V_{23}\right\vert =%
\sqrt{2}\frac{m_{s}}{m_{b}}$.

With respect to $|V_{12}|$ and the second Invariant in Eq. (\ref{ahoed}), we
compute $\widehat{h}_{u}=\frac{H_{u}^{-1}}{Tr[H_{u}^{-1}]}$, $\widehat{h}%
_{d}=\frac{H_{d}^{-1}}{Tr[H_{d}^{-1}]}$ and find in leading order 
\begin{equation}
\widehat{h}_{u,d}=\frac{1}{2}\left( 
\begin{array}{ccc}
1 & -1 & 0 \\ 
-1 & 1 & 0 \\ 
0 & 0 & 0
\end{array}
\right) +\frac{1}{2}\left( \frac{\beta }{\alpha }\right) _{u,d}\left( 
\begin{array}{ccc}
1 & 0 & -1 \\ 
0 & -1 & 1 \\ 
-1 & 1 & 0
\end{array}
\right) =\Delta _{o}+\widehat{\varepsilon }_{u,d}\widehat{A}_{u,d}
\label{ahoed1}
\end{equation}
which then leads to 
\begin{equation}
\left\vert \chi (\widehat{A})\right\vert =\left\vert \chi (\widehat{%
\varepsilon }_{d}\widehat{A}_{d})\right\vert =\frac{3}{4}\left( \frac{\beta
_{d}}{\alpha _{d}}\right) ^{2}  \label{usy12}
\end{equation}
Combining with Eqs. (\ref{ab}, \ref{ahoed})\ and with the results already
obtained for Eqs. (\ref{usyv23}, \ref{i2-usy0}), we find in leading order 
\begin{equation}
\left\vert V_{12}\right\vert ^{2}=\frac{m_{d}}{m_{s}}  \label{usy12new}
\end{equation}
which corresponds exactly to what was known for this USY ansatz.

Finally, putting together Eqs. (\ref{usyv23}, \ref{i2-usy0}, \ref{usy12new})
one obtains the correct USY approximate expression for $|V_{13}|=\frac{1}{%
\sqrt{2}}\frac{\sqrt{m_{d}m_{s}}}{m_{b}}$.

\subsection{Asymmetry in the NNI Weak Basis}

It has been shown \cite{nni}, that starting with arbitrary quark mass
matrices $M_{u}^{\circ }$, $M_{d}^{\circ }$, in the framework of the SM, it
is possible to make a WB transformation such that $M_{u}$, $M_{d}$ acquire
the Nearest Neighbour Interaction (NNI) form: 
\begin{equation}
M_{u}=c_{u}\left( 
\begin{array}{lll}
0 & a_{u} & 0 \\ 
\widehat{a}_{u} & 0 & b_{u} \\ 
0 & \widehat{b}_{u} & 1
\end{array}
\right) \qquad ;\qquad M_{d}=c_{d}\ K\cdot \left( 
\begin{array}{lll}
0 & a_{d} & 0 \\ 
\widehat{a}_{d} & 0 & b_{d} \\ 
0 & \widehat{b}_{d} & 1
\end{array}
\right)   \label{nni}
\end{equation}
where all $a_{u,d},\widehat{a}_{u,d},b_{u,d},\widehat{b}_{u,d},c_{u,d}$ are
real and the matrix $K=diag(1,e^{i\phi _{1}},e^{i\phi _{2}})$. In the limit
that $\widehat{a}_{u,d}=a_{u,d}$ and $\widehat{b}_{u,d}=b_{u,d}$, one
obtains the Fritzsch ansatz \cite{frit}, which has been eliminated by
experiment, namely by the large value of the top quark and the observed
value of $|V_{cb}|$.

In the following, we use our invariants to find out the minimal asymmetry
which is required in $M_{u}$, $M_{d}$, when written in the NNI basis, in
order to conform with experiment. Let us define the asymmetries 
\begin{equation}
\varepsilon _{u}\equiv \frac{\widehat{b}_{u}-b_{u}}{\widehat{b}_{u}+b_{u}}%
\qquad ;\qquad \varepsilon _{d}\equiv \frac{\widehat{b}_{d}-b_{d}}{\widehat{b%
}_{d}+b_{d}}  \label{eud}
\end{equation}
and the total asymmetry 
\begin{equation}
\varepsilon =\sqrt{\varepsilon _{u}^{2}+\varepsilon _{d}^{2}}  \label{aeud}
\end{equation}

Note that alignment and hierarchy of the quark mass matrices are guaranteed
in Eq. (\ref{nni}) by taking $(a,b)_{u,d}$, $(\widehat{a},\widehat{b})_{u,d}$
much smaller than 1. Computing the invariants associated to $h_{u}$ and $%
h_{d}$ as in Eq.(\ref{tr1}), and taking into account the hierarchy of the
quark mass matrices, one obtains in good approximation 
\begin{equation}
\left| a\right| ^{2}\ \left| \widehat{a}\right| ^{2}=\frac{\left|
m_{1}m_{2}\right| }{m_{3}^{2}}\qquad ;\qquad \left| b\right| ^{2}\ \left| 
\widehat{b}\right| ^{2}=\left( \frac{m_{2}}{m_{3}}\right) ^{2}  \label{ab1}
\end{equation}
Then, combining Eqs. (\ref{eud}, \ref{ab1}) we obtain: 
\begin{equation}
b_{u}^{2}=\frac{m_{c}}{m_{t}}\left( \frac{1-\varepsilon _{u}}{1+\varepsilon
_{u}}\right) \qquad ;\qquad b_{d}^{2}=\frac{m_{s}}{m_{b}}\left( \frac{%
1-\varepsilon _{d}}{1+\varepsilon _{d}}\right)   \label{bud}
\end{equation}

Now, computing $\chi (A)$ as in Eq. (\ref{diffh}) with $h_{u}$ and $h_{d} $
obtained from the NNI form in Eq. (\ref{nni}), and using Eq. (\ref{ab1}), we
get an expression which relates the experimental value for $\left|
V_{cb}\right| $ and the $\left| \chi (A)\right| $ as in Eq. (\ref{chia3}) in
terms of the parameters of the NNI form. We find in leading order 
\begin{equation}
b_{d}^{2}-2b_{u}b_{d}\cos (\phi )+b_{u}^{2}-b_{d}^{4}=\left| V_{cb}\right|
^{2}+2\left( \frac{m_{s}}{m_{b}}\right) ^{2}  \label{bud1}
\end{equation}
where $\phi =\phi _{1}-\phi _{2}$ is a complex phase resulting from the
diagonal matrix $K$ in Eq. (\ref{nni}). This expression is obtained taking
into account that $(a,\widehat{a})=O(\frac{\sqrt{\left| m_{1}m_{2}\right| }}{%
m_{3}}) $ and that $(b,\widehat{b})=O($ $\sqrt{\left| \frac{m_{2}}{m_{3}}%
\right| })$ as implied by Eqs. (\ref{ab1}, \ref{bud}).

From Eqs. (\ref{bud}, \ref{bud1}) we find that there is a connection between
the required asymmetries $\varepsilon _{u},\varepsilon _{d}$ of the up and
down quark sectors in order to conform to experiment. This connection can be
understood as follows. Take the case when $\phi =0$ and $\varepsilon _{d}=0$%
, then from the second relation in Eq. (\ref{bud}) it follows that $b_{d}=$ $%
\sqrt{\frac{m_{s}}{m_{b}}}$, but then the expression of Eq. (\ref{bud1})
forces also $b_{u}=$ $\sqrt{\frac{m_{s}}{m_{b}}}$ in leading order, and
therefore, from the first relation in Eq. (\ref{bud}), one gets $\varepsilon
_{u}\approx -1+2\left( \frac{m_{c}}{m_{t}}\right) /\left( \frac{m_{s}}{m_{b}}%
\right) $. Therefore, when the asymmetry in the down sector is small, the
required asymmetry in the up sector is large, and vice versa. It can be
readily verified that when $\phi \neq 0$, this result also holds. Indeed, by
eliminating from Eqs.(\ref{bud}, \ref{bud1}) both $b_{u}$ and $b_{d}$ one
finds $\varepsilon _{u}$ as a function of $\varepsilon _{d}$ and $\phi $ 
\begin{equation}
\varepsilon _{u}=\varepsilon _{u}(\varepsilon _{d},\phi )  \label{feud}
\end{equation}
and one then computes the total asymmetry $\varepsilon $ in Eq. (\ref{aeud}%
). This total asymmetry can thus be written as a function of $\varepsilon
_{d}$ and $\phi .$ One finds that it increases for all values of $\phi \neq 0
$, and it has a minimum for a certain value of $\varepsilon _{d}$ (and $\phi
=0$).

We have plotted the total asymmetry $\varepsilon $ in Fig. 1 as a function
of $\varepsilon _{d}$ for typical values of $\frac{m_{c}}{m_{t}}$ , $\frac{%
m_{s}}{m_{b}}$ and $\left\vert V_{cb}\right\vert $ at $M_{Z}$. As can be
seen from the plot, the minimal required total asymmetry is about $%
\varepsilon =0.2$, which indicates clearly that ans\"{a}tze, written in the
NNI basis, require quark mass matrices with a considerable amount of
asymmetry in order to conform to experiment. This finding agrees with the
result previously obtained \cite{emagus} by explicitly diagonalizing the
quark mass matrices written in the NNI basis.

\begin{figure}[!t]
\begin{center}
\includegraphics[width=10cm,height=7cm]{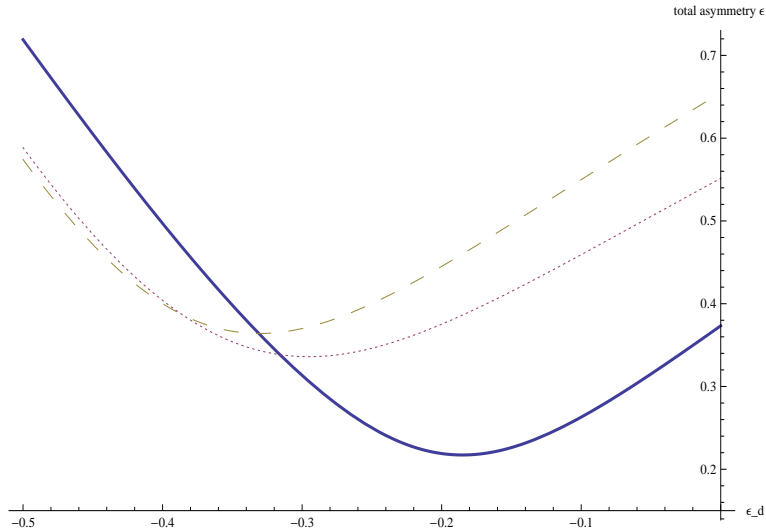}
\end{center}
\caption{Total required asymmetry $\varepsilon$ as a function of the down
quark mass matrix asymmetry $\varepsilon_d $. The full line is for the
values of $m_s = 60 MeV, \phi=0$, the tiny dashed line for values of $%
m_s=100\ MeV, \phi=0$ and large dashed line for values of $m_s=80\ MeV,
\phi=0.35$. For all curves, we took the values for $m_b=3.0\ GeV $, $%
m_c=680\ MeV$, $m_t=181\ GeV$ and $|V_{cb}|=0.037$ at $M_Z$. }
\label{fig:asym}
\end{figure}

\section{Leptons}

The hierarchy of lepton masses may also be expressed in terms of invariants
of $H_{l},H_{\nu }$. For the charged leptons one may use essentially the
same invariants as the quarks. For the neutrinos, the invariant 
\begin{equation}
R_{1}\equiv \frac{4\chi [H_{\nu }]}{Tr[H_{\nu }]^{2}}  \label{hier2n}
\end{equation}
may distinguish normal hierarchy corresponding to $R_{1}\ll 1$ from inverted
hierarchy ($R_{1}=1$) and degeneracy ($R_{1}=\frac{4}{3}$). The invariant 
\begin{equation}
R_{2}\equiv \frac{3Tr[H_{\nu }]\ Det[H_{\nu }]}{\left( \chi [H_{\nu
}]\right) ^{2}}  \label{hier3n}
\end{equation}
may also be used to distinguish inverted hierarchy when $R_{2}$ is small
from degeneracy when $R_{2}=1$. Furthermore, for normal hierarchy, it can
distinguish the cases when one of the two smaller masses is much small then
the other one, $R_{2}\ll 1$, or the case when these two small masses are of
the same order. In this case $R_{2}\ $\ is of order one$.$ Thus, we have 
\begin{equation}
\begin{array}{lllllll}
Normal_{1} &  & Normal_{2} &  & Inverted &  & Degenerate \\ 
R_{1}\ll 1 &  & R_{1}\ll 1 &  & R_{1}=1 &  & R_{1}=\frac{4}{3} \\ 
&  &  &  &  &  &  \\ 
R_{2}\ll 1 &  & R_{2}=O(1) &  & R_{2}\ll 1 &  & R_{2}=1
\end{array}
\label{tab}
\end{equation}

\subsection{Normal Hierarchy}

For definiteness, let us assume that neutrinos have normal hierarchy, with $%
m_{1}^{2}\ll m_{2}^{2}\ll m_{3}^{2}$. Next, we show that one can use a set
of WB invariants of the lepton sector in order to fix the leptonic mixing
matrix. In the limit, $m_{e}=0$, one obtains: 
\begin{equation}
I_{\nu _{2}}\equiv 1-\frac{Tr[H_{l}]\ Tr[H_{l}H_{\nu }]-Tr[H_{l}^{2}H_{\nu }]%
}{\chi [H_{l}]\ Tr[H_{\nu }]}=\frac{\frac{m_{1}^{2}}{m_{3}^{2}}\
|V_{12}|^{2}+|V_{13}|^{2}}{1+\frac{m_{1}^{2}}{m_{2}^{2}}}  \label{limit-v13}
\end{equation}
The experimental data show that the leptonic mixing matrix is close to the
tribimaximal mixing \cite{tribi}. In particular, it is shown that $%
|V_{13}|^{2}\equiv |U_{e3}|^{2}$ is small. From Eq. (\ref{limit-v13}), it
follows that this can be guaranteed by having $I_{\nu _{2}}\ll 1$. On the
other hand, the associated invariant $I_{\nu _{2}}^{\prime }$, obtained by
interchanging $H_{l}$ and $H_{\nu }$ yields in the limit $m_{1}=0$: 
\begin{equation}
I_{\nu _{2}}^{\prime }\equiv 1-\frac{Tr[H_{\nu }]\ Tr[H_{\nu
}H_{l}]-Tr[H_{\nu }^{2}H_{l}]}{\chi [H_{\nu }]\ Tr[H_{l}]}=\frac{\frac{%
m_{e}^{2}}{m_{\tau }^{2}}\ |V_{21}|^{2}+|V_{31}|^{2}}{1+\frac{m_{e}^{2}}{%
m_{\tau }^{2}}}  \label{v13n}
\end{equation}
By putting $I_{\nu _{2}}^{\prime }\approx \frac{1}{6}$, one has $%
|V_{31}|^{2}\approx \frac{1}{6}$ as in the tribimaximal mixing. Then,
computing the leptonic invariant equivalent to that of Eqs. (\ref{i10}, \ref
{chia3}), one obtains: 
\begin{equation}
\left| \chi \left[ A_{\nu }\right]\right| =\left| V_{23}\right| ^{2}+\left|
V_{13}\right| ^{2}+O\left( \left( \frac{m_{2}}{m_{3}}\right) ^{4}\right)
\label{chian}
\end{equation}
which must be near $\frac{1}{2}$ for tri-bimaximal mixing. With $I_{\nu
_{2}} $, $I_{\nu _{2}}^{\prime }$ and $\chi \left[ A_{\nu }\right] $ and the
unitarity of $V^{PMNS}$ one can ensure that the mixing is near to the
tri-bimaximal mixing: if $|V_{13}|^{2}$ is small and $\left| V_{23}\right|
^{2}$ is near $\frac{1}{2}$, then $\left| V_{33}\right| ^{2}$ must also be
near to $\frac{1}{2}$. Since $\left| V_{31}\right| ^{2}$ is near to $\frac{1%
}{6}$, it follows that $\left| V_{32}\right| ^{2}$ is near to $\frac{1}{3}$
and then $|V_{12}|^{2}$ and $|V_{11}|^{2}$ are near to $\frac{4}{6}$ and $%
\frac{1}{3}$ respectively.

We have thus shown how to obtain the observed pattern of leptonic mixing
through a set of invariant conditions.

\section{Conclusions}

We pointed out that the use of weak-basis invariants can avoid the well
known redundancy of free parameters in the flavour structure of mass
matrices. These invariants are specially useful when one opts for a
bottom-up approach to the study of the flavour structure of Yukawa couplings
and fermion mass matrices. In particular, we have shown that the pattern of
fermion mixing both in the quark and lepton sectors can be expressed in
terms of relations only involving weak-basis invariants. We have also
pointed out the observed alignment of the up and down quark mass matrices in
flavour space can also be guaranteed through a weak-basis invariant
condition. It was emphasized that in the context of the SM, the above
alignment in no way follows automatically from the Yukawa couplings
structure, since $Y_{u}$ and $Y_{d}$\ are independent. On the other hand,
this alignment may arise naturally, e.g. in left-right symmetric theories or
in $SO(10)$, where $Y_{u}$ and $Y_{d}$\ may be approximately proportional to
each other.

In summary, WB invariants may play an important r\^{o}le in a systematic
search for patterns of fermion mass matrices consistent with experiment and
may thus help to uncover a possible flavour symmetry chosen by nature.

\section{\protect\bigskip Acknowledgments}

This work was partially supported by Funda\c{c}\~{a}o para a Ci\^{e}ncia e a
Tecnologia (FCT, Portugal) through the projects CERN/FP/11638/2010,
PTDC/FIS/098188/2008 and CFTP-FCT Unit 777 which are partially funded
through POCTI (FEDER) and by Marie Curie Initial Training Network "UNILHC"
PITN-GA-2009-237920.


\begin{thebibliography}{99}
\bibitem{mix}  Particle Data Group, J. Phys. G \textbf{37}, 075021 (2010).

\bibitem{ckm}  N. Cabibbo, Phys. Rev. Lett. \textbf{10} (1963) 531; M.
Kobayashi and T. Maskawa, Prog. Theor. Phys. \textbf{49} (1973) 652.

\bibitem{bsm}  F.J. Botella, G.C. Branco, M. Nebot and M.N. Rebelo, Nucl.
Phys. B \textbf{725} (2005) 155.

\bibitem{neut0}  For a recent global fit, see Thomas Schwetz, Mariam Tortola
and J.W.F. Valle, New J. Phys. \textbf{13} (2011) 109401.

\bibitem{principles}  J. Bernab\'{e}u, G.C. Branco and M. Gronau, Phys.
Lett. B \textbf{169} (1986), 243.

\bibitem{cp-quark}  G.C. Branco and M.~N.~Rebelo, Phys.\ Lett.\ B \textbf{173%
} (1986) 313; M. Gronau, A. Kfir and R. Loewry, Phys.\ Lett.\ B \textbf{56}
(1986) 1538; G.C. Branco and V.A. Kostelecky, Phys. Rev. D \textbf{39 (}%
1989) 2075; F.J. Botella, M. Nebot and O. Vives, JHEP \textbf{0601} (2006)
106.

\bibitem{cp-lepton}  G.C. Branco, L. Lavoura and M.N. Rebelo, Phys.\ 
Lett.\
B \textbf{180} (1986) 264; G.C. Branco, M.N. Rebelo and J.W.F. Valle,
Phys.\ Lett.\ B \textbf{225} (1989) 385; F. del Aguila and J.A.
Aguilar-Saavedra, Phys.\ Lett.\ B \textbf{386} (1996) 241; A. Pilaftsis,
Phys. Rev. D \textbf{56 (}1997) 5431; G.C. Branco, M.N. Rebelo and J.I.
Silva-Marcos, Phys. Rev. Lett. \textbf{82} (1999) 683; G.C. Branco, T.
Morozumi, B.M. Nobre and M.N. Rebelo, \ Nucl. Phys.\ B \textbf{617} 
(2001)
475; S. Davidson and R. Kitano, JHEP \textbf{0403 (}2004) 020; H.K. Dreiner,
J.S. Kim, O. Lebedev and M. Thormeier, Phys. Rev. D \textbf{76 (}2007)
015006; P.F. Harrison and W.G. Scott, Phys.\ Lett.\ B \textbf{628} (2005) 93;

\bibitem{cp-higgs}  L. Lavoura and J.P. Silva, Phys. Rev. D \textbf{50 (}%
1994) 4619; F.J. Botella and J.P. Silva, Phys, Rev. D \textbf{51 (}1995)
3870; G.C. Branco, M.N. Rebelo and J.I. Silva-Marcos Phys.\ Lett.\ B 
\textbf{614} (2005) 187; J.F. Gunion and H.E. Haber, Rev. D \textbf{72 (}%
2005) 095002;

\bibitem{previously}  G.C. Branco and Dan-di Wu, Phys.\ Lett.\ B \textbf{205}
(1988) 353.

\bibitem{usy}  G.~C.~Branco, J.~I.~Silva-Marcos and M.~N.~Rebelo, Phys.\
Lett.\ B \textbf{237} (1990) 446.

\bibitem{usy1}  Paul M. Fishbane, Peter Kaus, Phys. Rev. D \textbf{49}
(1994) 4780; J.~Kalinowski and M.~Olechowski, Phys.\ Lett.\ B \textbf{251}
(1990) 584.

\bibitem{usy2}  G.C. Branco, J.I. Silva-Marcos, Phys. Lett. B \textbf{359}
(1995) 166; G.C. Branco, D. Emmanuel-Costa, J.I. Silva-Marcos, Phys. Rev. D 
\textbf{56 (}1997) 107; J.~I.~Silva-Marcos, Phys. Lett. B \textbf{443}
(1998) 276; G.C. Branco, H.R. Cola{\c{c}}o Ferreira, A.G. Hessler and 
J.I. Silva-Marcos, arXiv:1101.5808.

\bibitem{nni}  G.C. Branco, L. Lavoura and F. Mota, Phys. Rev. D \textbf{39 (%
}1989) 3443.

\bibitem{frit}  H. Fritzsch, Phys. Lett. B \textbf{73} (1978) 317.

\bibitem{emagus}  G.C. Branco, D. Emmanuel-Costa and C. Sim\~{o}es, Phys.
Lett. B \textbf{690} (2010) 62.

\bibitem{tribi}  P.F. Harrison, D.H. Perkins and W.G. Scott, Phys. Lett. B 
\textbf{530} (2002) 167.
\end{thebibliography}
\end{document}